\documentclass[aps,prb,amsmath,amssymb,reprint,superscriptaddress]{revtex4-1}

\usepackage{epstopdf}

\usepackage{graphicx}
\usepackage{amsmath}
\usepackage{dcolumn}
\usepackage{bm}
\usepackage{bbold}
\usepackage{epstopdf}
\usepackage{amssymb}
\usepackage{mathrsfs}
\usepackage{amsfonts}
\usepackage{color}
\usepackage{mathtools}
\usepackage{braket}
\usepackage{tabularx}
\usepackage{setspace}
\usepackage{dcolumn}
\usepackage{bm}

\usepackage{float}
\usepackage{array}
\usepackage{caption}
\usepackage{subcaption}
\newcolumntype{L}[1]{>{\raggedright\let\newline\\\arraybackslash\hspace{0pt}}m{#1}}
\newcolumntype{C}[1]{>{\centering\let\newline\\\arraybackslash\hspace{0pt}}m{#1}}
\newcolumntype{R}[1]{>{\raggedleft\let\newline\\\arraybackslash\hspace{0pt}}m{#1}}

\begin{document}

\preprint{AIP/123-QED}

\title[]{Low-temperature microwave properties of biaxial YAlO$_3$} 

\author{N. C. Carvalho}
\email[]{nataliaccar@gmail.com}
\affiliation{School of Physics, The University of Western Australia, 35 Stirling Hwy, 6009 Crawley, Western Australia.}
\affiliation{ARC Centre of Excellence for Engineered Quantum Systems (EQuS), 35 Stirling Hwy, 6009 Crawley, Western Australia.}

\author{M. Goryachev}%
\affiliation{School of Physics, The University of Western Australia, 35 Stirling Hwy, 6009 Crawley, Western Australia.}
\affiliation{ARC Centre of Excellence for Engineered Quantum Systems (EQuS), 35 Stirling Hwy, 6009 Crawley, Western Australia.}

\author{J. Krupka}%
\affiliation{Instytut Mikroelektroniki i Optoelektroniki PW, Koszykowa 75, 00-662 Warsaw, Poland}

\author{P. Bushev}%
\affiliation{Experimentalphysik, Universit\"{a}t des Saarlandes, D-66123 Saarbr\"{u}cken, Germany}

\author{M. E. Tobar}
\affiliation{School of Physics, The University of Western Australia, 35 Stirling Hwy, 6009 Crawley, Western Australia.}
\affiliation{ARC Centre of Excellence for Engineered Quantum Systems (EQuS), 35 Stirling Hwy, 6009 Crawley, Western Australia.}%

\date{\today}

\begin{abstract}
Low-loss crystals with defects due to paramagnetic or rare earth impurity ions is a major area of investigation for quantum hybrid systems at both optical and microwave frequencies. In this work we  examine the single crystal yttrium aluminium perovskite, YAlO$_3$ using the Whispering Gallery Mode Technique. Multiple resonant microwave modes were measured from room temperature to 20 mK allowing precise characterization of the permittivity tensor at microwave frequencies. We show that it is biaxial and characterize the tensor as a function of temperature with estimated uncertainties below 0.26\%. Electron spin resonance spectroscopy was also performed at 20 mK, with new transitions identified with Zero-Field splittings of 16.72 and 9.92 GHz. Spin-photon couplings of order 4.2 and 8.4 MHz were observed for residual levels of concentration, which are stronger than the photon cavity linewidths of 116 kHz but the same order of the linewidths of the discovered spin transitions.
%
\end{abstract}

\pacs{Valid PACS appear here}
\maketitle


\section{Introduction}

The yttrium aluminum perovskite (YAP) is an inorganic material well known for its interesting mechanical and chemical properties. Its high light yield and good resolution have made it suitable for a variety of applications in optical physics, ranging from solid state lasers \cite{liuhan2015} and scintillators \cite{yasuda2000} to medical apparatus \cite{randazzo2008} and recording media \cite{loutts1998}.

This rare-earth (RE) aluminate has also demonstrated very good performance in the microwave range, with a relative permittivity value of order 16 and low loss\cite{konaka1991}, which are valuable for the design of dielectric resonators. Also, due its high heat conductivity, it has found use as a substrate material for thin films of high temperature superconductors with a wide range for microwave component applications, such as use in microstrip lines \cite{cho1999}.

Recently, experimentalists have shown that RE electron spin defects in YAP have great potential for the development of crucial components for quantum computers and quantum communication devices \cite{lovric2012}.  In fact, strong coupling has been demonstrated in circuit QED experiments using erbium doped YAP, Er$^{3+}$:YAlO$_{3} $\cite{tkalvcec2014} at microwave frequencies, with the aim to implement quantum memories using hybrid architecture. However, the microwave properties of YAP are not well characterized and such experiments would greatly benefit from a rigorous characterization of the material properties, especially at dilution refrigerator temperatures.

The cavity-resonator method has been implemented previously\cite{konaka1991} to measure the relative permittivity and dielectric losses of YAP. This work presented a single permittivity component at a frequency of 7.767 GHz equal to 16 at room temperature. This determined value decreased no more than 4\% when the crystal was cooled down to 20 K. However, YAP is a biaxial crystal with orthorhombic symmetry; which means it has a diagonal permittivity tensor with three distinct elements as in Eq. \ref{Eq1} \cite{jellison2011, petit2010}. Therefore, full determination of its permittivity tensor components is desirable and will allow more accurate design in applications using this material.

\begin{equation}
	\begin{pmatrix}
		\epsilon_x &0 & 0 \\
  		0 & \epsilon_y & 0 \\
  		0 & 0 & \epsilon_z 
 	\end{pmatrix}
	\label{Eq1}
\end{equation}

For those reasons, this work presents for the first time a thorough characterization of real permittivity tensor of YAP using the Split-Post Dielectric Resonator (SPDR) and Whispering Gallery Mode (WGM) techniques. Temperature dependence of the tensor components and the crystal losses down to 20 mK were also investigated and temperature coefficients of permittivity (TCP) were calculated. Following this we performed spectroscopy of impurity ions in the sample using the WGM technique \cite{benmessai2013,farr2013,goryachev20142, goryachev2015} at 20 mK. New impurity ion transitions were measured with Zero-Field splittings of  16.72 and 9.92 GHz. Their correspondent interaction with the microwave photonic modes were also analyzed with respective coupling strengths determined.

\section{Dielectric Characterisation}
\renewcommand{\arraystretch}{1.75}

\begin{table}
		\caption{YAP real permittivity temperature dependence.}
		\label{tab1}
				\begin{tabular}[c]{ C{2.7cm}  C{1.5cm}  C{1.5cm}  C{1.5cm}}  \hline \hline
Temperature (K)  &  $\epsilon_x$ & $\epsilon_y$ & $\epsilon_z$\\ \hline 
0.020	&16.316	&15.535	&15.128\\
1	&16.316	&15.535	&15.128\\
4	&16.321	&15.531	&15.126\\
10	&16.321	&15.531	&15.126\\
23	&16.323	&15.530	&15.126\\
29	&16.324	&15.530	&15.126\\
39	&16.327	&15.531	&15.128\\
50	&16.328	&15.535	&15.133\\
58	&16.333	&15.537	&15.139\\
65	&16.341	&15.538	&15.145\\
69	&16.342	&15.542	&15.149\\
295	&16.582	&15.997	&15.673\\  \hline \hline
				\end{tabular}
\end{table}
\begin{figure}[h!]
\includegraphics[width=0.42\textwidth]{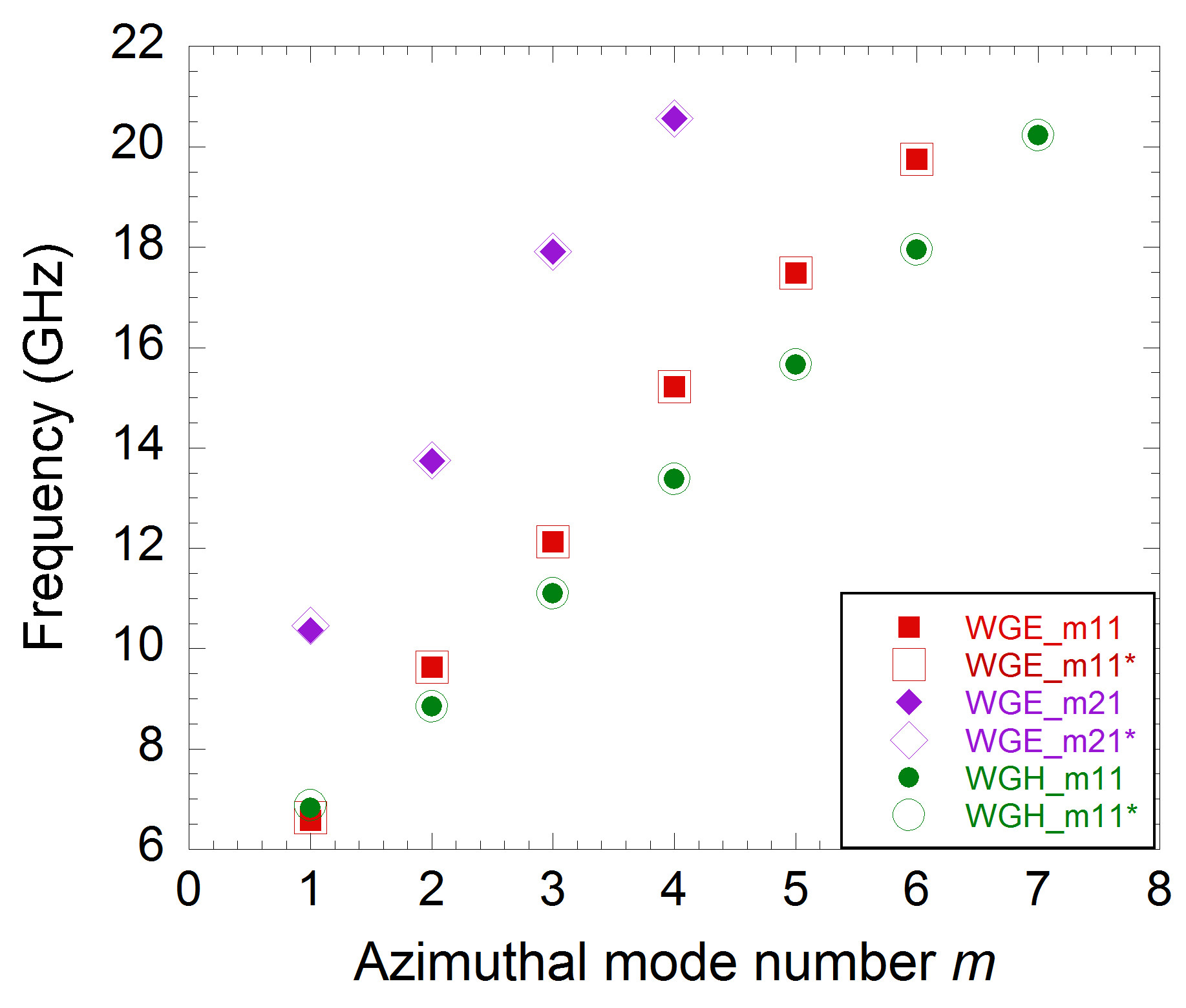}
\caption{\label{fig1}Room temperature measurement of the mode frequency versus azimuthal mode number (\textit{m}).}
\end{figure} 
Through the SPDR technique \cite{krupka2001, lefloch2014} the three components of the real part of the complex permittivity tensor of the YAP were initially determined only at room temperature and with low precision. To perform this measurement three flat laminar crystal samples of 0.52 mm thickness with 1\% variation between samples were used. Such samples had the remaining dimensions equal to 10 mm $\times$ 10 mm and each square face perpendicular to one of the YAP principal axes. The SPDR technique uses the transverse electric mode to probe the permittivity of the two crystal axes perpendicular to the thickness direction. In this way, an initial estimate of the permittivity tensor may be provided. For the YAP crystal the permittivity components were measured as $\epsilon_x = 16.4\pm 0.2$,  $\epsilon_y  = 15.7\pm 0.2$ and  $\epsilon_z  = 15.3\pm 0.2$. 

In order to have a more accurate measure of the permittivity tensor for a bulk YAP sample, we then implemented the more precise WGM technique \cite{tobar2001, krupka1999}, allowing us to refine the preliminary results. The WGM method uses a cylindrical bulk sample placed within a cylindrical metallic cavity. Multiple electromagnetic modes are then excited within the dielectrically loaded cavity and the WGMs families are selected to perform the permittivity characterization. 

A cylindrical YAP sample of $10.44 \pm 0.01$ mm height and diameter equal to $12.00 \pm 0.01$ mm made by Scientific Materials Corporation was used. It had a concentric hole $1.550 \pm 0.001$ mm diameter parallel to its longitudinal axis, also parallel to the crystal crystalographic a-axis. The sample was placed into a copper cavity and supported by a Teflon holder to position it at near the center of the cavity (cavity design in detail in \cite{carvalho2015}). Two electromagnetic probes were inserted into the cavity and connected to the input/output signal of a Vector Network Analyzer (VNA) operating in transmission mode. 

The fundamental WGM photonic mode families in the YAP crystal were recorded in a frequency range from 6 to 20 GHz at room temperature as shown in Fig. \ref{fig1}. In this frequency range typically the real part of permittivity of dielectric crystals have negligible frequency dependence \cite{fsc}. So to calculate the permittivity tensor components a Finite Element Method (FEM) model of the resonator was implemented, first using the initial estimates of the permittivity tensor components from the SPDR technique. The permittivity values were then incremented around this value with steps of $\Delta \epsilon_i = 0.001, i = x, y, z$ until the best match between the experimental (shown in Fig. \ref{fig1}) and simulated modes was reached.

The real permittivity tensor at room temperature for YAP was obtained as shown at the last line of Table \ref{tab1}. The discrepancy between the experimental mode frequencies and the simulated data were typically 10 MHz on average for the fundamental WGM families, with the uncertainties mainly due the irregularities of the crystal dimensions. To perform a temperature dependent determination of the real permittivity, two WGMs of different polarization must be selected \cite{krupka1999, krupka19992, carvalho2015}. We chose the WGH$_{311}$ and WGE$_{511}$ modes as they had well distinguished doublet modes required to determine the biaxial nature and were of high enough order to ensure enough mode confinement in the crystal\cite{carvalho2015}. 

\begin{figure}[h!]
\includegraphics[width=0.38\textwidth]{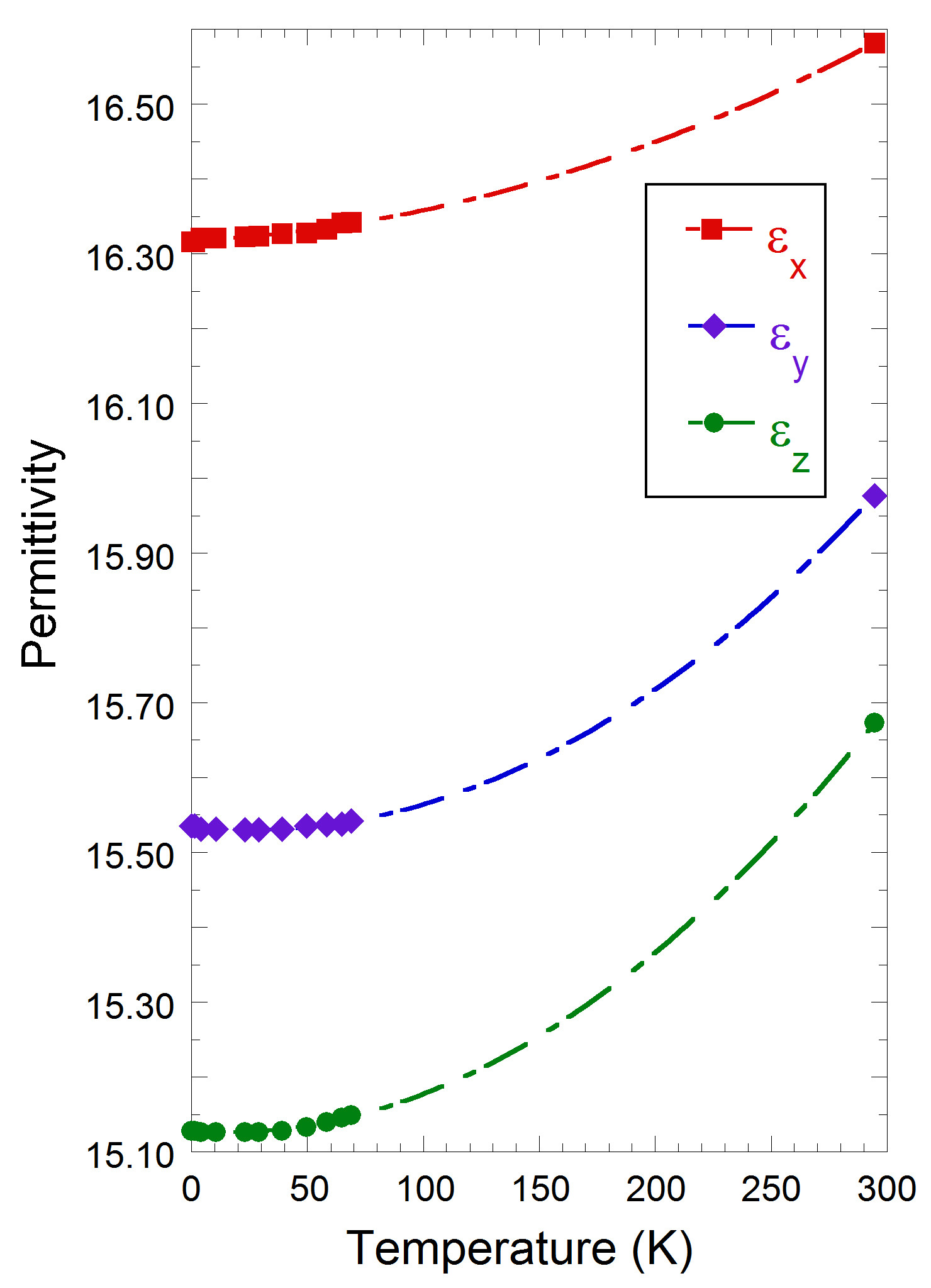}
\caption{\label{fig2} Measurement of the YAP real permittivity tensor components as a function of temperature.}
\end{figure}
\begin{figure}[h!]
\includegraphics[width=0.45\textwidth]{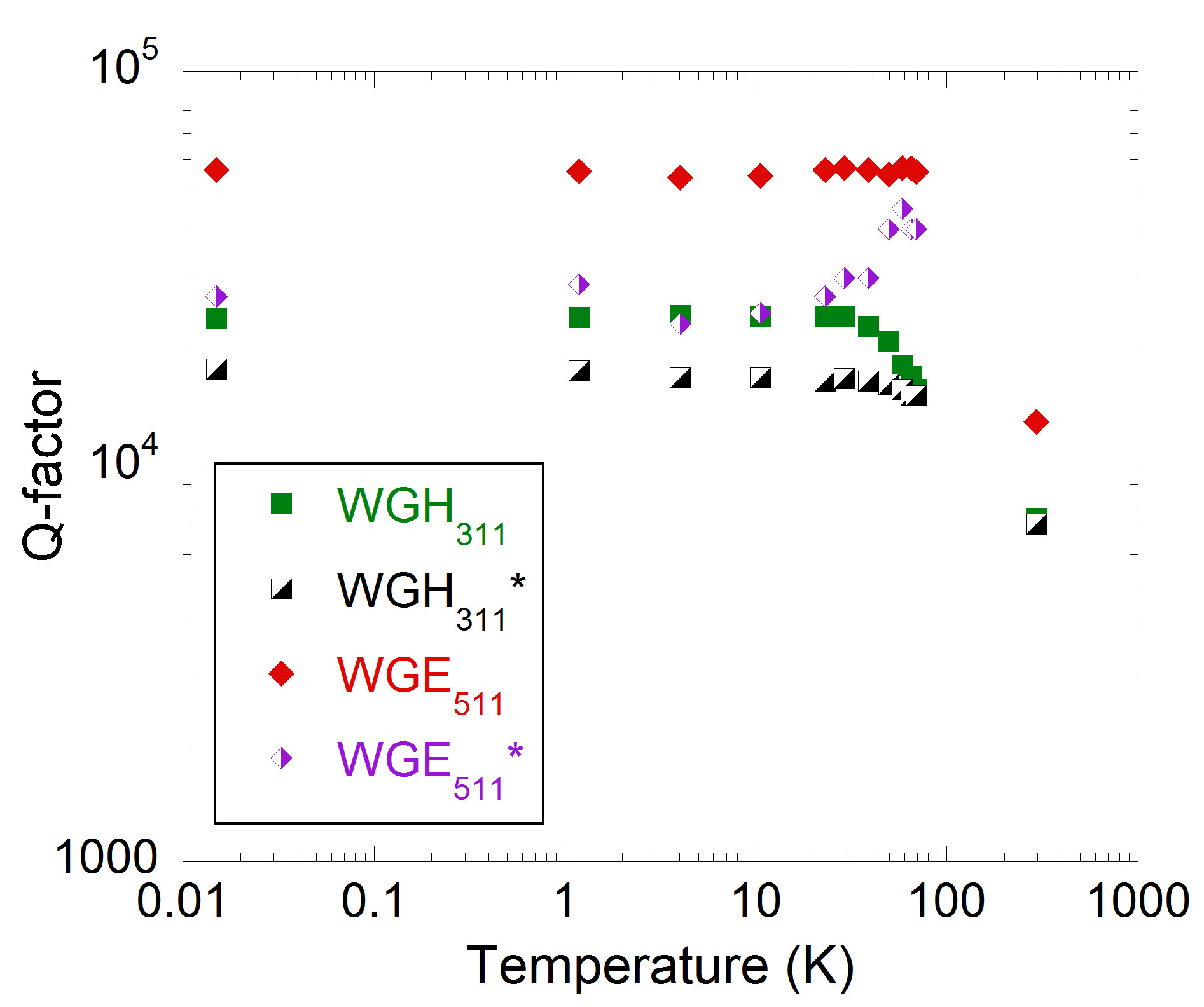}
\caption{\label{fig3}Measured temperature dependence of the electrical Q-factor of the modes WGE$_{311}$ and WGH$_{511}$ and their respective doublets pairs (marked with an asterisk).}
\end{figure}
\begin{figure}[h!]
\includegraphics[width=0.42\textwidth]{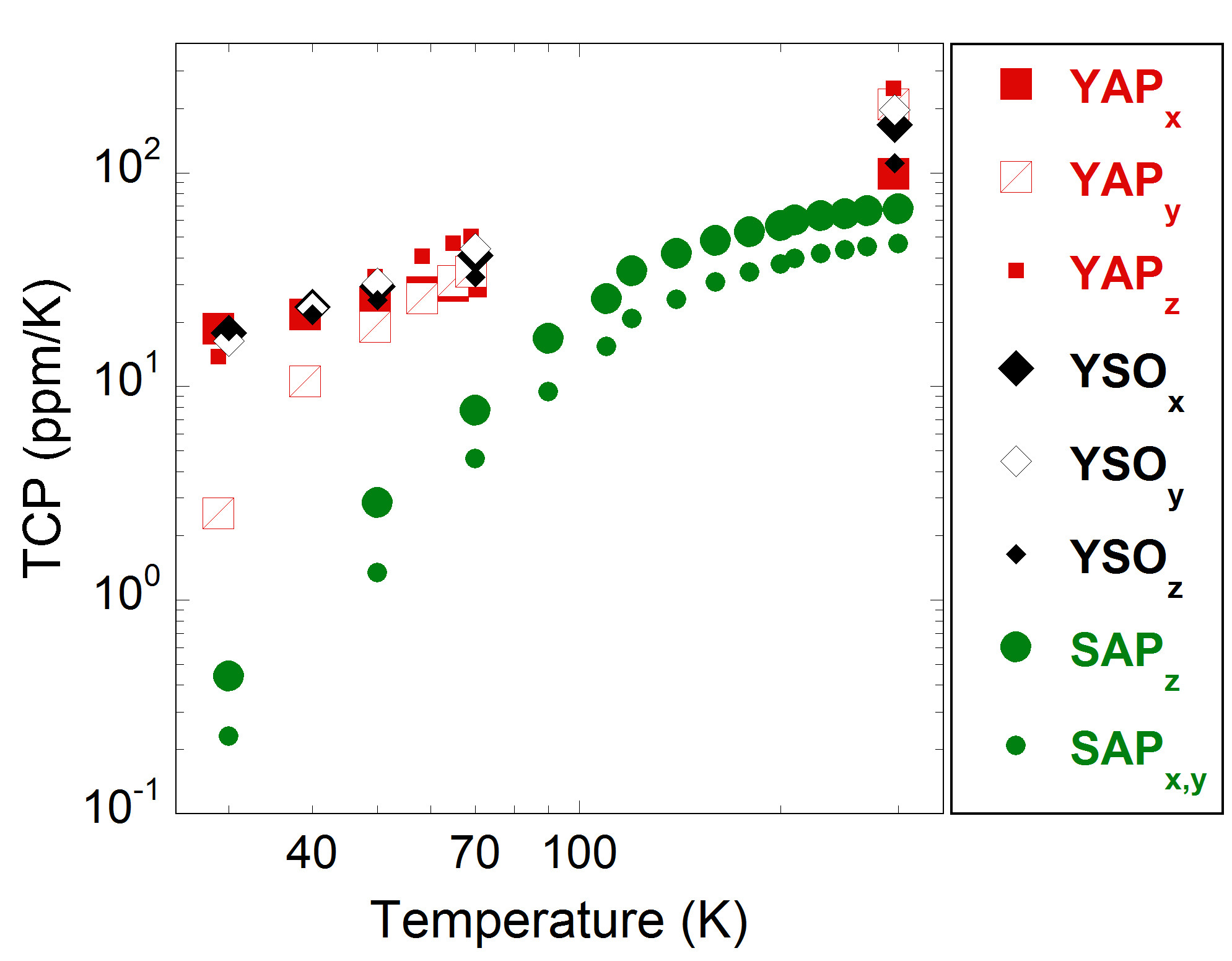}
\caption{\label{fig4} TCP data of the biaxial YAP, YSO \cite{carvalho2015} and uniaxial sapphire (SAP) \cite{krupka1999} crystal permittivity elements.}
\end{figure}
The cavity containing the sample was placed into a Blue Fors LD-250 dilution refrigerator and cooled down to 20 mK. All the selected mode frequencies and electrical quality factors (Q-factors) were tracked and recorded over the warming up process. The FEM model was implemented for an additional 11 different temperatures measurements to determine the three components of permittivity as a function of temperature. The thermal contraction of the copper cavity and the sample were taken into consideration, whereas the Teflon contribution was determined to be negligible from the FEM model.

Table \ref{tab1} shows the calculated permittivity tensor over the measured temperature range. They are also presented in the Fig. \ref{fig2}, alongside with their respective second order polynomial fit. These results are consistent with the results published by \cite{konaka1991} and show no significant variation below 40 K. Under 1 K, small frequency shifts with temperature are observed, which are below the precision of the permittivity determination. This effect has been well documented in cryogenic sapphire resonators and was shown to be due to residual paramagnetic impurity defects within the crystal lattice \cite{hartnett}. This outcome has been used to annul the Temperature Coefficient of Frequency (TCF) of high-Q microwave WGMs, by operating at the temperature where the temperature coefficient of magnetic susceptibility of the impurity ion cancels the shift due to the TCP \cite{kb}. This is essential for the realization of the ultra stable cryogenic sapphire oscillator, which attains exquisite frequency stability of order parts in $10^{16}$ \cite{nand2013}. Similar paramagnetic defects in sapphire have also been used to create high stability masers \cite{kb,maser}, precision frequency conversion \cite{fc} and the generation of high stability frequency combs due to four wave mixing \cite{fwm}.

Fig. \ref{fig3}  presents how the crystal losses change in cryogenic temperatures. The Q-factor of the resonance modes was measured at the same temperatures the permittivity components were probed. The results show the losses decrease by about three to four times from 300 K to 40 K and do not vary much more for colder temperatures.

The TCP was also calculated and in Fig. \ref{fig4} the YAP coefficients are compared to other two widely used single crystals. The data points are only presented for temperatures above 30 K, away from effects due to residual paramagnetic impurities.

An immediate analysis shows that YAP has a stronger temperature dependence of permittivity than sapphire, however sapphire is not a good host for RE elements \cite{koechner1996}, which is an undesirable feature for several applications. Comparing to the Yttrium Orthosilicate (YSO), YAP's TCP does not have significant difference, but still has the advantage of having relatively lower magnetic anisotropy, which make it more suitable for operation in superconducting cavities and circuits \cite{ asatryan1997}. 

\section{Electron Spin Resonance Spectroscopy at 20 mK}

The evidence of paramagnetic impurity defects was originally observable from the frequency temperature dependence of the microwave modes at low temperature. Thus, due to the potential of the YAP crystal for microwave and optical quantum information applications, we performed an Electron Spin Resonance (ESR) Spectroscopy close to the lowest temperature of our measurements, where spins in the lattice may be coupled strongly with photons and manipulated with qubits \cite{tkalvcec2014}. Moreover, recently iron group impurity ion spin states were discovered in single crystal YSO with strong coupling to whispering gallery photons\cite{goryachev2014}, and in this work we confirm similar spin ensembles in YAP.

\begin{figure}[h!]
\includegraphics[width=0.495\textwidth]{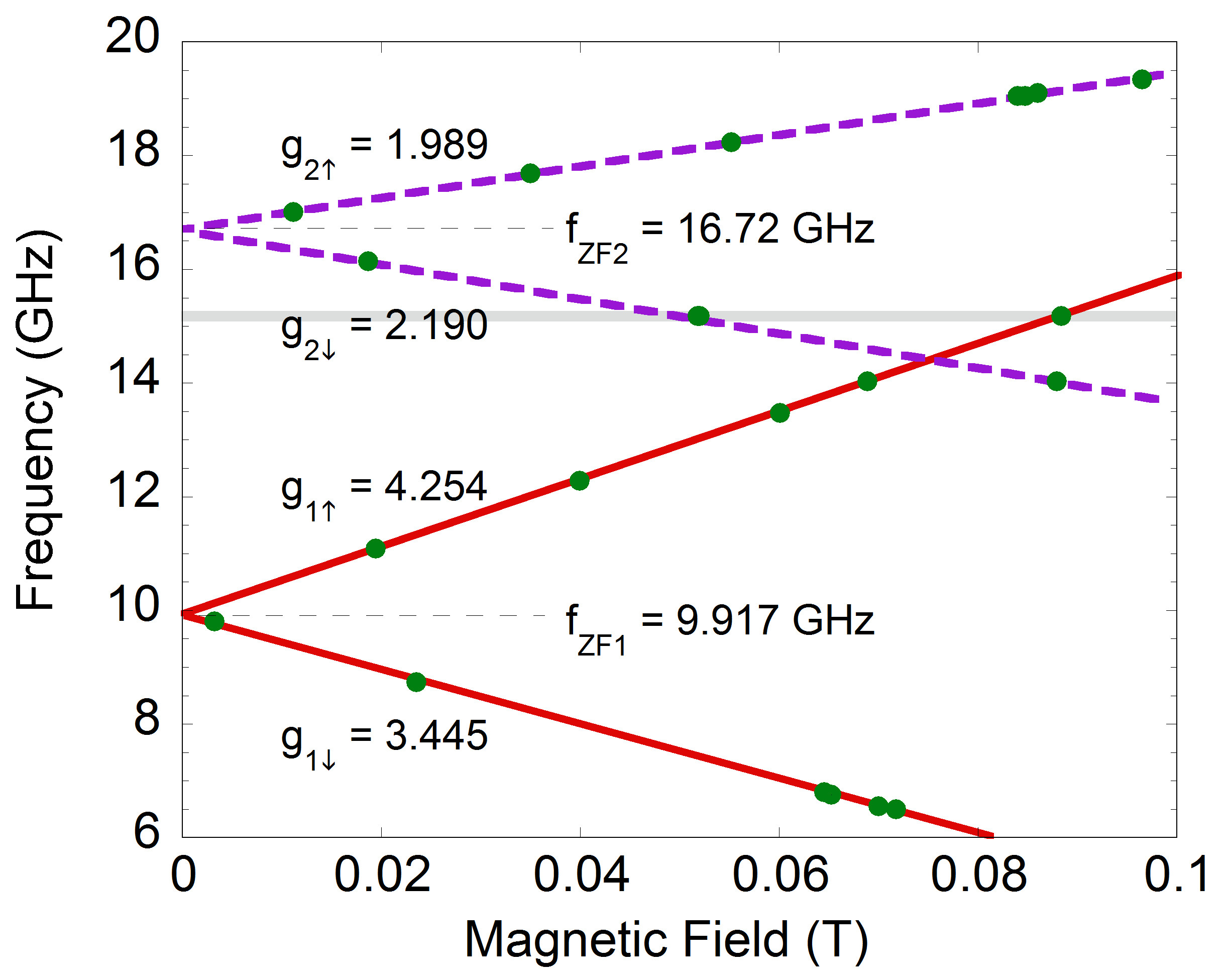}
\caption{\label{fig5} Energy level splitting of two ionic impurity transitions with applied magnetic field. Each dot corresponds to an Avoided Level Crossing (ALC), the horizontal shaded line is representing the WGM at 15.18 GHz and the angled solid and broken lines are showing the Zeeman effect of the $f_{ZF1}$ and $f_{ZF2}$ Zero-Field splittings, respectively. The effective g-factors of the transitions are also included.}
\end{figure}

\begin{figure}[h!]
\includegraphics[width=0.495\textwidth]{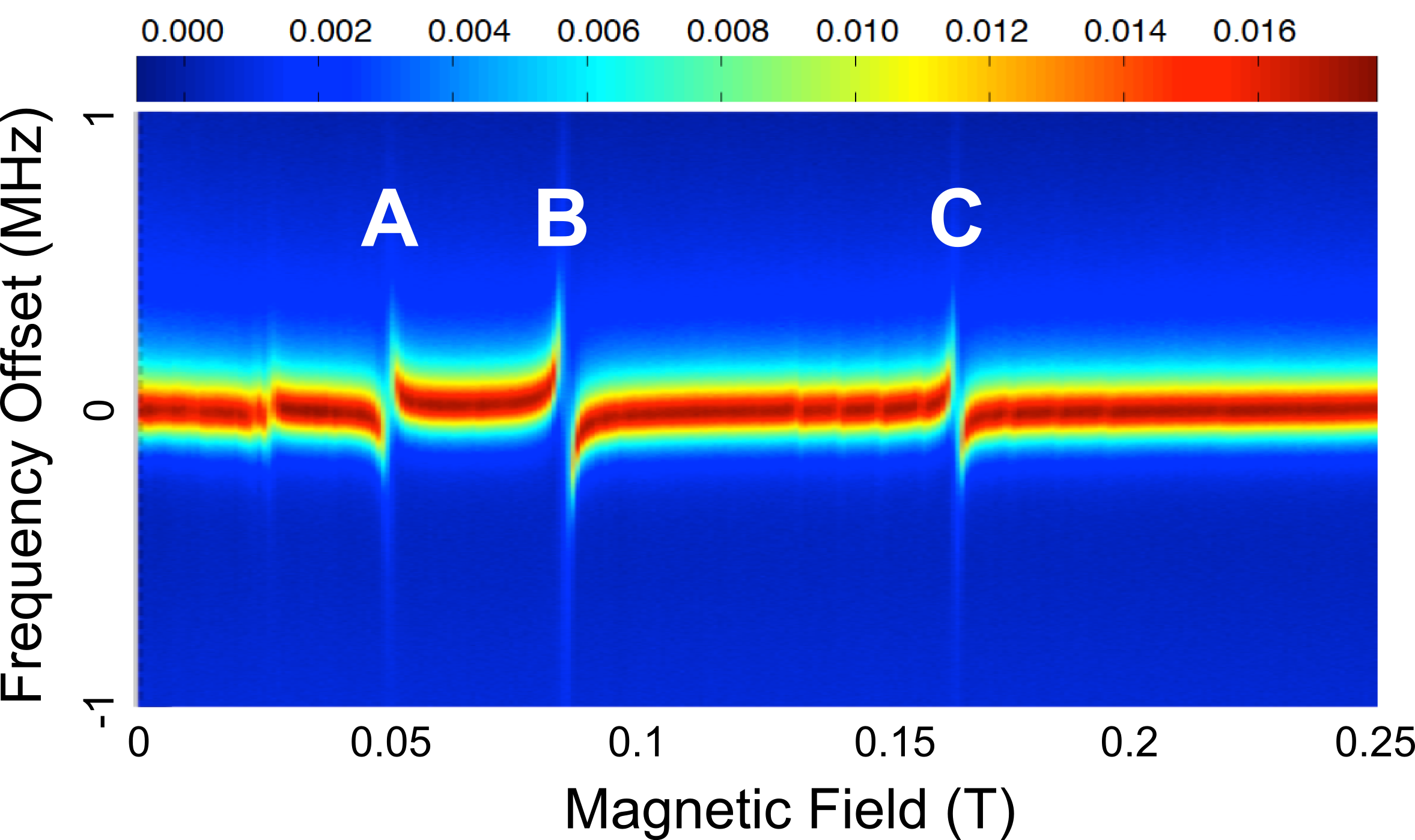}
\caption{\label{fig6} Magnetic field dependence of the cavity transmission from 0 to 0.25 T. Three interactions under study between the 15.18 GHz photonic mode and the crystal spins ensembles are labeled by the letters A, B and C. More weakly coupled spins transitions are present below 0.03 T, however these were not investigated any further due to poor signal-to-noise ratio, which was not good enough to be seen in the other WGMs at the spread of frequencies shown in Fig. \ref{fig5}.}
\end{figure}

The experiment consisted of placing the copper cavity containing the YAP cylinder into a superconducting magnet in such a way that a DC magnetic field would be parallel to the crystal axis (more details of the experimental apparatus can be found in \cite{farr2013}). Transmission measurements identified several WGMs generated in the crystal between 6 and 20 GHz and at 20 mK in temperature. These modes were then monitored in a narrow frequency span and at low input power while the magnet performed a sweep from 0 to 0.1 T. 

The data collected from the VNA showed the signatures of several interactions between spin states and photonic modes as indicated by Avoided Level Crossings (ALCs), i.e., the normal mode splitting resultant from the hybridization of two harmonic oscillators. Fig. \ref{fig5} presents a map where the relevant ALCs are represented by dots and are connected by lines to emphasize the Zeeman effect. The two indicated Zero-Field splittings ($f_{ZF1}$ and $f_{ZF2}$) at 16.720 and 9.917 GHz, demonstrate the presence of two ionic impurity transitions in the crystal lattice. They are characterized by the indicated effective g-factors, $g_L$, which are related to the nature of unintentional dopants and are calculated from $g_L = \frac{df}{dB}(\frac{2 \pi \hbar}{\mu_B})$, where $df/dB$ is the slope of line fitting the transition and $\mu_B$ is the Bohr magneton.

Particularly interesting due to apparently higher spin-photon couplings, the WGM at 15.18 GHz (highlighted by the shaded horizontal line in Fig. \ref{fig5}) was subject to further investigation. Fig. \ref{fig6} shows the transmission spectroscopy plot of the spin-photon interactions when the magnetic field is changed from 0 to 0.25 T. In this plot, the ALCs A and B are explained by the hybridization of the photonic mode with the $g_{2\downarrow}$ and the $g_{1\uparrow}$ spins, respectively. In contrast, the mode splitting C indicates the coupling with an RE impurity, responsible for the additional splitting caused by the hyperfine (HF) transitions between states. 

\begin{figure*}
\includegraphics[width=0.95\textwidth]{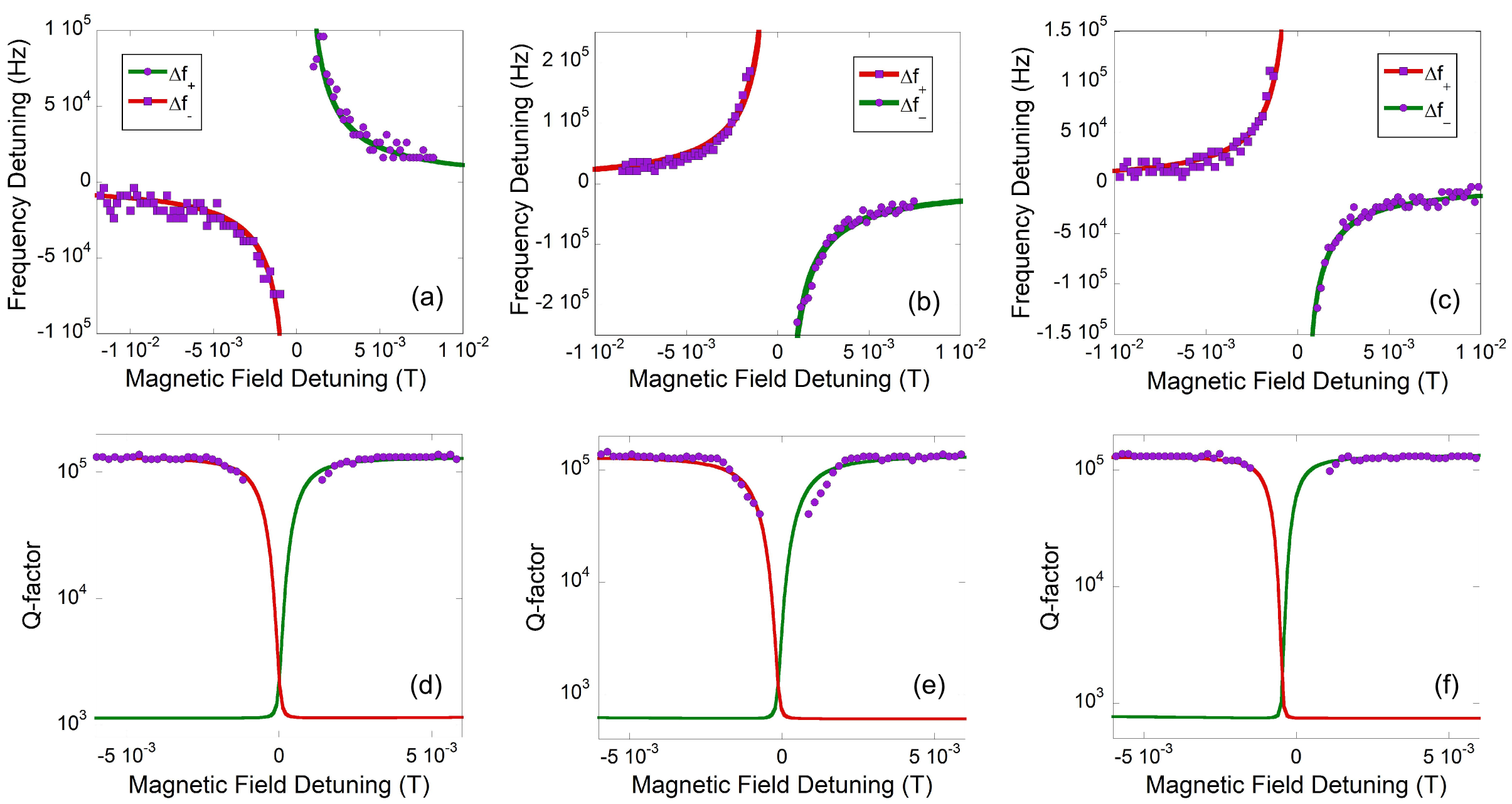}
\caption{\label{fig7} Frequency shift as a function of the magnetic field: (a) A, (b) B and  (c) C normal mode splittings due the interactions between the spins ensembles and the 15.18 GHz WGM. $\Delta f_+$ and $\Delta f_-$ correspond to the two coupled modes; (d), (e) and (f) shows Q-factors of the coupled modes as a function of the magnetic field and correspond to the interactions A, B and C, respectively. The magnetic field zero-detunings are 0.516 T in (a) and (d), 0.88 T in (b) and (e) and 0.169 T in (c) and (f).}
\end{figure*}

\setlength{\tabcolsep}{11pt}
\renewcommand{\arraystretch}{1.75}
\begin{table*}[]
\centering
\caption{ESR data: spin-photon interactions of the 15.18 GHz WGM shown in Figures \ref{fig6} and \ref{fig7}. }
\label{tab2}
\begin{tabular}{ c c c c c c c }
\hline \hline
 & B (Tesla) & $\delta_{spins}/2$ (MHz) & $\Gamma_{WGM}/2$ (MHz) & g (MHz) & $Q_{spins}$ & $Q_{WGM}$ \\
\hline
A & 0.516     & 6.7                      & 0.058                   & 2.1     & 1130        & $1.3 \times 10^5$ \\ 
B & 0.088     & 12.5                     & 0.058                   & 4.2     & 606         & $1.3 \times 10^5$ \\ 
C & 0.169     & 10.3                     & 0.058                   & 3.4     & 740         & $1.3 \times 10^5$ \\
\hline \hline
\end{tabular}
\end{table*}

In reference \cite{tkalvcec2014}, the resonant photon interaction with the electron spin and corresponding HF structure of the $^{167}Er$ isotope with a nuclear spin of I = 7/2 was rigorously researched. In Fig. \ref{fig6} a similar prominent ALC is recognised, due to the transition between $m_s = \pm 1/2$ spin states, surrounded by eight HF transitions. In contrast, the transitions associated with the A and B ALCs have never before been identified, and are thus of potential interest for application in hybrid quantum architectures.

Many experiments that characterise the hybridisation of spin ensembles and photons use an input-output model that calculates the transmission spectra at the point of the mode crossing\cite{Schuster2010,Dinz2011,Sandner2012}. However in our experiments, due to the fact that the coupling between the resonator photons and spin ensemble is smaller than the inhomogeneous spin linewidth, means that the effective Q-factor of the spin ensemble dominates, and the transmission spectra at the point of the mode crossing is severely degraded and becomes an inaccurate technique to characterise the system due to the very poor signal to noise ratio. This technique is based on a two-mode coupled harmonic oscillators model, an older paper shows how the parameters derived from reflection coefficient or transmission spectra of a coupled mode system can also be successfully obtained by solving the characteristic eigenvalue equations\cite{Tobar1991mtt, tobar1993}. This is a very good technique in this limit, as we can use the full range of the magnetic field and the normal mode frequency and linewidth (or effective Q-factor) variations with sufficient signal-to-noise ratio to fit the relevant parameters during the interaction. Thus, to determine the properties of the 15.18 GHz WGM interacting with the measured spins ensembles, the ALCs were fitted using a similar model.

This model assumes two representative LCR oscillators (labeled with subscripts 1 and 2) coupled reactively through the magnetic field (mutual inductance) with a dimensionless coupling coefficient of $2\Delta _g$ \cite{Tobar1991mtt, tobar1993}. The calculated characteristic equation for the coupled mode system is written in Eq. \ref{Eq2}, ignoring all terms of order $1/Q^2$. This is a very good approximation for any system of relatively high Q-factor (i.e. above 10). 

\begin{equation}
\left. \begin{array}{ll}
\displaystyle \omega ^4+\omega ^3 \left(\frac{\omega _1}{Q_1}+\frac{\omega_2}{Q_2}\right)+\omega ^2 \left(\omega _1^2+\omega _2^2\right)+ \\
\displaystyle \omega  \left(\frac{\omega _2 \omega _1^2}{Q_2}+\frac{\omega _2^2 \omega _1}{Q_1}\right)+\omega _1^2 \omega _2^2-2 \Delta _g \omega _1^2 \omega _2^2
\end{array} \right. 
 \label{Eq2}
\end{equation}

To solve for the normal mode frequencies, and hence coupling between the normal modes, all Q-values were set to infinity and the normal mode frequencies, $\omega_+=2\pi f_+$ and $\omega_-=2\pi f_-$, were solved for by setting the characteristic equation to zero. The only unknown to determine was the coupling strengths, which in Hz is given by $g = \omega_{WGM}/2\pi\times\Delta _g$, and was determined by fitting the experimental data (assigning subscript 1 to the WGM and subscript 2 to the spin ensemble). Results of the fits  for the three photon-spin interactions are shown in Fig. \ref{fig7} (a)-(c), with the numeric values of $g$ listed in Table \ref{tab2}. 

The hybridized spin-photon system clearly shows a coupling greater than the WGM photon linewidth. However, due the inhomogeneous broadened spin ensembles in the interaction, it was difficult to estimate the spin linewidth (or effective spin Q-factor) as the cavity coupling was reduced to below the sensitivity of our experiment, and it could not be determined if we were in the strong coupling regime. Thus, to solve for the spin linewidths (or effective spin Q-factor) we used a similar technique as developed in reference \cite{tobar1995} for a coupled n-mode system of mechanical oscillators. To do this we write the characteristic equation for the normal mode system as:

\begin{equation}
\left. \begin{array}{ll}
\displaystyle \omega ^4+\omega ^3 \left(\frac{\omega _-}{Q_-}+\frac{\omega _+}{Q_+}\right)+\omega ^2\left(\omega _-^2+\omega _+^2\right)+ \\
\displaystyle  \omega\left(\frac{\omega _+ \omega _-^2}{Q_+}+\frac{\omega _+^2 \omega_-}{Q_-}\right)+\omega _-^2\omega _+^2
\end{array} \right.    
 \label{Eq3}
\end{equation}

Then, we equate odd powers of $\omega$ between Eq. \ref{Eq2} and Eq. \ref{Eq3} (note, equating even coefficients verifies the same normal mode frequencies as already determined). Subsequently the following matrix equation may be derived to solve for normal mode Q-factors:

\begin{equation}
\left(
\begin{array}{c}
 \frac{\omega _+}{Q_+} \\
 \frac{\omega _-}{Q_-} \\
\end{array}
\right)=
\left(
\begin{array}{cc}
 1 & 1 \\
 \omega _-^2 & \omega _+^2 \\
\end{array}
\right)^{-1}
\left(
\begin{array}{cc}
 1 & 1 \\
 \omega _2^2 & \omega _1^2 \\
\end{array}
\right)
\left(
\begin{array}{c}
 \frac{\omega _1}{Q_1} \\
 \frac{\omega _2}{Q_2} \\
\end{array}
\right)
 \label{Eq4}
\end{equation}

The photonic Q-factor, $Q_{WGM}$ ($Q_1$) was easily measured at zero field and by measuring the normal mode Q-factors during the mode hybridization, the effective Q-factor of the spin ensemble, $Q_{spins}$ ($Q_2$), may be estimated by fitting the data to the matrix equation above (given by Eq. \ref{Eq4}). The data along with the fitting curves are shown in Fig. \ref{fig7} (a)-(f) and the extracted parameters are shown in Table \ref{tab2}.

Although the coupling strengths of the three interactions exceed the photonic half linewidths ($\Gamma_{WGM}$), the spins half linewidths ($\delta_{spin}$) surpassed the couplings by a factor of three, which means the inhomogeneous spin linewidth prevented the strong coupling regime ($g > \Gamma_{WGM}/2, \delta_{spin}/2$) to be reached. The collective coupling strength, nevertheless, is related to the spins concentration, $n$, by $g = g_L \mu_B \sqrt{\frac{\mu_0 \omega_0 n \xi}{4 \hbar}}$ \cite{bushev2011}, where $\omega_0$ and $\xi$ are the frequency of the WGM and the sample transverse magnetic filling factor, respectively. For the 15.18 GHz resonance, $\xi$ is about 84\%, corresponding to the simulated $WGH_{213}$ mode structure. Thus, the estimated average spin concentration is on the order of 10$^{14}$ cm$^{-3}$ and increasing the concentration of impurities in the crystal by intentional doping would allow strong coupling between the photonic mode and the spins ensembles.

\section{Conclusions}

In conclusion, this work presented an investigation of the YAP microwave properties. In the first part, a precise measurement of the three non-zero components of the YAP real permittivity tensor was performed and the values of $\epsilon_x = 16.316$, $\epsilon_y = 15.535$ and $\epsilon_z = 15.128$ were obtained with the sample at 20 mK. Mode frequencies predicted from the found permittivity tensor differ on average by only 10 MHz from the measured data, with the uncertainties mainly due the imprecision in the determination of the sample's dimensions. The crystal losses were also assessed through the measurement of the Q-factors of four photonic modes and a study of the temperature dependence of these properties were performed between room and sub-Kelvin temperatures, including the measurement of the temperature coefficient of permittivity. In the second part,  ESR spectroscopy of the same sample was performed at 20 mK and allowed the identification of two new electronic spin impurity transitions with Zero-Field splitting at 16.72 and 9.92 GHz and spin-photon couplings strengths of $2g = 4.2$ and $8.4$ MHz, respectively, exceeding the WGM linewidth of 116 kHz. Spin ensembles coupled to a microwave resonator is a promising architecture for the reversible transfer of quantum states between systems in the GHz range \cite{bertet}, in the sense the presented results represent a valuable information for the design of experiments aiming to implement the use of YAP crystals in hybrid quantum systems.

\begin{acknowledgments}

This research is supported by the Australian Research Council, CE110001013 and by the Conselho Nacional de Desenvolvimento Cient\'ifico e Tecnol\'ogico (CNPq – Brazil). The authors would like to thank Warrick Farr and Nikita Kostylev with some help with data acquisition and processing.

\end{acknowledgments}

\end{document}